\documentclass[preprintnumbers,showpacs,amsmath,amssymb,floatfix,twocolumn,prd,superscriptaddress,nofootinbib]{revtex4}
\usepackage{graphicx}
\usepackage{epsfig}
\usepackage{bm}
\usepackage{amsfonts}

\begin{document}

\title{Generalized second law of thermodynamics for a phantom energy accreting BTZ black hole}

\author{Mubasher Jamil}
\email{mjamil@camp.nust.edu.pk} \affiliation{Center for Advanced
Mathematics and Physics,\\ National University of Sciences and
Technology, Rawalpindi, 46000, Pakistan}

\author{M. Akbar}
\email{makbar@camp.nust.edu.pk} \affiliation{Center for Advanced Mathematics and Physics,\\
National University of Sciences and Technology, Rawalpindi, 46000,
Pakistan}

\begin{abstract}

\textbf{Abstract:} In this paper, we have studied the accretion of
phantom energy on a (2+1)-dimensional stationary
Banados-Teitelboim-Zanelli (BTZ) black hole. It has already been
shown by Babichev et al that for the accretion of phantom energy
onto a Schwarzschild black hole, the mass of black hole would
decrease and the rate of change of mass would be dependent on the
mass of the black hole. However, in the case of (2+1)-dimensional
BTZ black hole, the mass evolution due to phantom accretion is
independent of the mass of the black hole and is dependent only on
the pressure and density of the phantom energy. We also study the
generalized second law of thermodynamics at the event horizon and
construct a condition that puts an lower bound on the pressure of
the phantom energy.
\end{abstract}


 \maketitle
\newpage
\section{Introduction}
It has been found by various astronomical and cosmological
observations \cite{perl} that our universe is currently in the phase
of accelerated expansion. In the framework of Einstein's gravity,
this accelerated expansion has been explained by the presence of a
`cosmological constant' bearing negative pressure which results in
the stretching of the spacetime \cite{pad}. Many other theoretical
models have been presented to explain the present accelerated
expansion of the universe including based on homogeneous and time
dependent scalar field like the quintessence \cite{essence},
Chaplygin gas \cite{chaplygin} and phantom energy \cite{phantom}, to
name a few. The phantom energy is characterized by the equation of
state $p=\omega\rho$, with $\omega<-1$. It possesses some weird
properties: the cosmological parameters like energy density and
scale factor become infinite in a finite time; all gravitationally
bound objects lose mass with the accretion of phantom energy; the
fabric of spacetime is torn apart at the big rip; and that it
violates the standard relativistic energy conditions. The
astrophysical data coming from the microwave background radiation
categorically favors the phantom energy \cite{cald}. Motivated from
the dark energy models, we model phantom energy by an ideal fluid
with negative pressure.

The accretion of dark energy onto a black hole has been studied by
many authors \cite{jamil} after the seminal work of Babichev et al
\cite{babi} who have shown that the mass of the black hole will
decrease with time when we consider the accretion of phantom energy.
In the Einstein theory of gravity, the accretion of the phantom
energy onto Schwarzschild black hole and evaporation of primordial
black hole has been discussed \cite{babi,jamil1}. It will be
interesting to investigate the accretion dynamics in low and higher
dimensional gravities. It is also important to investigate accretion
dynamics in the extended theories of gravity.

In this paper we investigate the accretion of exotic phantom energy
onto a static uncharged 3-dimensional BTZ black hole. We will show
that the expression of the evolution of BTZ black hole mass is
independent of its mass and dependents only on the energy density
and pressure of the phantom energy. It is well-known that the
horizon area of the black hole decreases with the accretion of
phantom energy, hence it is essential to study the generalized
second law of thermodynamics (GSL) in this case \cite{GSL}. We show
that the validity of GSL in the present model yields an lower bound
on the phantom energy pressure. We also demonstrate that the first
law of thermodynamics holds in the present construction.

The plan of the paper is as follows: In second section we model the
accretion of phantom energy onto three dimensional BTZ black hole.
In third section, we study the GSL for BTZ black hole. Finally we
conclude our results.

\section{Model of accretion}
Consider the field equations for a (2+1)-dimensional spacetime with
a negative cosmological constant $\Lambda$
\begin{equation}
G_{ab}+\Lambda g_{ab}=\pi T_{ab}, \ \ (a,b=0,1,2)
\end{equation}
where $G_{ab}$ is the Einstein tensor in (2+1)-dimension while
$T_{ab}$ is the stress energy tensor of the matter field. The units
are chosen such that $c=1$ and $G_3=1/8$. Considering the
stress-energy tensor to be vacuum, one can obtain the following
spherically symmetric metric, a (2+1)-dimensional BTZ black hole
\cite{ban}
\begin{equation}
ds^2=-f(r)dt^2+\frac{dr^2}{f(r)}+r^2d\phi^2,
\end{equation}
where $f(r)=-M+r^2/l^2$, $M$ is the dimensionless mass of the black
hole and $l^2=-1/\Lambda$, is a positive constant. The coefficient
$g_{00}$ is termed as the lapse function. The event horizon of the
BTZ black hole is obtained by setting $f(r)=0$, which turns out,
$r_e= l\sqrt{M}$. Also we have $\sqrt{|g|}=r$, where $g$ is the
determinant of the metric. To analyze the accretion of phantom
energy onto the BTZ black hole, we here employ the formalism from
the work by Babichev et al \cite{babi}. The stress energy momentum
tensor representing the phantom energy is the perfect fluid
\begin{equation}
T^{ab}=(\rho+p)u^a u^b+pg^{ab},
\end{equation}
Here $\rho$ and $p$ are the energy density and pressure of the
phantom energy while $u^a=(u^0,u^1,0)$ is the velocity three vector
of the fluid flow. Also $u^1=u$ is the radial velocity of the flow
while the third component $u^2$ is zero due to spherical symmetry of
the BTZ black hole. There are two important equations of motion in
our model: one which controls the conservation of mass flux is
$J^{a}_{;a}=0$, where $J^a$ is the current density and the other
that controls the energy flux $T^{a}_{0;a}=0$, across the horizon.
Since the black hole is stationary, the only component of stress
energy tensor of interest is $T^{01}$. Thus the equation of energy
conservation $T^{0a}_{;a}=0$ is
\begin{equation}
ur(\rho+p)\sqrt{f(r)+u^2}=C_1,
\end{equation}
where $C_1$ is an integration constant. Since the flow is inwards
the black hole therefore $u<0$. Also the projection of the energy
momentum conservation along the velocity three vector $u_a
T^{ab}_{;b}=0$ (the energy flux equation) is
\begin{equation}
ur\exp\Big[\int\limits_{\rho_\infty}^{\rho_h}\frac{d\rho}{\rho+p}\Big]=-A_1.
\end{equation}
Here $A_1$ is a constant and the associated minus sign is taken for
convenience. Also $\rho_h$ and $\rho_\infty$ are the energy
densities of phantom energy at the BTZ horizon and at infinity
respectively. From Eqs. (4) and (5), we obtain
\begin{equation}
(\rho+p)\sqrt{f(r)+u^2}\exp\Big[-\int\limits_{\rho_\infty}^{\rho_h}\frac{d\rho}{\rho+p}\Big]=C_2,
\end{equation}
where $C_2=-C_1/A_1=\rho_\infty+p(\rho_\infty)$. The rate of change in
the mass of black hole $\dot M=-2\pi r T^1_0$, is given by
\begin{equation}
dM=2\pi A_1(\rho_\infty+p_\infty)dt.
\end{equation}
Note that $\rho_\infty+p_\infty<0$ (violation of null energy
condition) leads to decrease in the mass of the black hole.
Moreover, the above expression is also independent of mass contrary
to the Schwarzschild black hole and the Reissner-Nordstr\"{o}m black
hole \cite{jamil}. Further, the last equation is valid for any
general $\rho$ and $p$ violating the null energy condition, thus we
can write
\begin{equation}
dM=2\pi A_1(\rho+p)dt.
\end{equation}

\section{Critical Accretion}
We are interested only in those solutions that pass through the
critical point as these correspond to the material falling into the
black hole with monotonically increasing speed. The falling fluid
can exhibit variety of behaviors near the critical point of
accretion, close to the compact object. The equation of mass flux or
the continuity equation $J^a_{;a}=0$ is
\begin{equation}
\rho u r=k_1.
\end{equation}
Here $k_1$ is integration constant. From Eqs. (4) and (9), we have
\begin{equation}
\Big(\frac{\rho+p}{\rho}\Big)^2\Big(f(r)+u^2
\Big)=\Big(\frac{C_1}{k_1}\Big)^2=C_3.
\end{equation}
Taking differentials of (9) and (10) and after simplification, we
obtain
\begin{equation}
\frac{du}{u}\Big[-V^2 +\frac{u^2}{f(r)+u^2} \Big]+\frac{dr}{r}\Big[
-V^2+ \frac{r^2}{l^2\Big(f(r)+u^2\Big)} \Big]=0.
\end{equation}
Here
\begin{equation}
V^2\equiv\frac{d\text{ln}(\rho+p)}{d\text{ln}\rho}-1,
\end{equation}
From (11) if one or the other bracket factor is zero, one gets a
turnaround point corresponding double-valued solution in either $r$
or $u$. The only solution that passes through a critical point is
feasible. The feasible solution will correspond to material falling
into the object with monotonically increasing velocity. The critical
point is obtained by taking the both bracketed factors in Eq. (11)
to be zero. This will give us the critical points of accretion. We
obtain
\begin{eqnarray}
V_c^2&=&\frac{r_c^2}{(f(r_c)+u_c^2)l^2},\\
V_c^2&=&\frac{u_c^2}{f(r_c)+u_c^2}.
\end{eqnarray}
Above the subscript $c$ refers to the critical quantity. On
comparing Eqs. (13) and (14), we get
\begin{equation}
u_c^2=\frac{r_c^2}{l^2},\ \ V_c^2=\frac{u_c^2}{-M+2u_c^2}.
\end{equation}
Here $u_c$ is the critical speed of flow at the critical points
which we determine below. For physically acceptable solution, we
require $V_c^2>0$, hence we get the following restrictions on speeds
and the location of the critical points
\begin{equation}
u_c^2>\frac{M}{2},\ \ r_c^2>\frac{r_+^2}{2}.
\end{equation}

\section{ Generalized second law of thermodynamics and BTZ black hole}
In this section we will discuss the thermodynamic of phantom energy
accretion that crosses the event horizon of BTZ black hole. Let us
first write the BTZ metric in the form
\begin{equation}
ds^{2} = h_{mn}dx^{m}dx^{n}+r^{2}d\phi^{2}, \ \ \ m,n=0,1
\end{equation}
where $h_{mn}= \text{diag}(-f(r), 1/f(r))$, is a 2-dimensional
metric. From the condition of normalized velocities $u^{a}u_{a} =
-1$, one can obtain the relations
\begin{equation}
u^{0} = f(r)^{-1}\sqrt{f(r) + u^{2}}, ~~~ u_{0} = -\sqrt{f(r) +
u^{2}}.
\end{equation}
The components of stress energy tensor are $T^{00}=f(r)^{-1}[(\rho +
p)(\frac{f(r)+u^{2}}{f(r)})-p]$, and $T^{11}=(\rho + p)u^2+f(r)p$.
These two components help us in calculating the work density which
is defined by $W=-\frac{1}{2}T^{mn}h_{mn}$ \cite{cai}. In our case
it comes out
\begin{equation}
W=\frac{1}{2}(\rho - p).
\end{equation}
The energy supply vector is defined by
\begin{equation}
\Psi_{n}=T^{m}_{n}\partial_{m}r + W\partial_{n}r.
\end{equation}
The components of the energy supply vector are $\Psi_{0}=
T^{1}_{0}=-u(\rho+p)\sqrt{f(r)+u^2}$, and $\Psi_{1}=
T^{1}_{1}+W=(\rho+p)\Big(\frac{1}{2}+\frac{u^2}{f(r)}\Big)$. The
change of energy across the apparent horizon is determined through
$-dE\equiv-A\Psi$, where $\Psi=\Psi_0dt+\Psi_1dr$. The energy
crossing the event horizon of the BTZ black hole is given by
\begin{equation}
dE=4\pi r_eu^2(\rho+p)dt.
\end{equation}
Assuming $E=M$ and comparing Eqs. (8) and (21), we can determine the
value of constant $A_1=2u^2l\sqrt{M}$.

The entropy of BTZ black hole is
\begin{equation}
S_h=4\pi r_e.
\end{equation}
It can be shown easily that the thermal quantities, change of
phantom energy $dE$, horizon entropy $S_h$ and horizon temperature
$T_h$ satisfy the first law $dE=T_hdS_h$, of thermodynamics. After
differentiation of last equation w.r.t. $t$, and using Eq. (8), we
have
\begin{equation}
\dot{S}_h=8\pi^2l^2u^2(\rho+p).
\end{equation}
 Since all the parameters are positive in the above equation (23) except
that $\rho+p<0$, it shows that the second law of thermodynamics is
violated i.e. $\dot{S}_h<0$, as a result of accretion of phantom
energy on a BTZ black hole.\\ Now we proceed to the generalized
second law of thermodynamics (GSL). It is defined by
\begin{equation}
\dot{S}_{tot}=\dot{S}_h+\dot{S}_{ph}\geq0.
\end{equation}
In other words, the sum of the rate of change of entropies of black
hole horizon and phantom energy must be positive. We consider event
horizon of the BTZ black hole as a boundary of thermal system and
the total matter energy within the event horizon is the mass of the
BTZ black hole. We also assume that the horizon temperature is in
equilibrium with the temperature of the matter-energy enclosed by
the event horizon, i.e. $T_h=T_{ph}=T$, where $T_{ph}$ is the
temperature of the phantom energy. Similar assumptions for the
temperatures $T_h$ and  $T_{ph}$ has been studied in \cite{davies}.
We know that the Einstein field equations satisfy first law of
thermodynamics $T_hdS_h=pdA+dE$, at the event horizon \cite{azad}.
We also assume that the matter-energy enclosed by the event horizon
of BTZ black hole also satisfy the first law of thermodynamics given
by
\begin{equation}
T_{ph}dS_{ph}=pdA+dE.
\end{equation}
 Here the horizon
temperature is given by
\begin{equation}
T_h=\left.\frac{f^\prime(r)}{4\pi}\right\vert_{r=r_e}
=\frac{\sqrt{M}}{2\pi l}.
\end{equation}
In this paper, we are assuming that $T_h=T_{ph}=T$. Therefore Eq.
(24) gives
\begin{equation}
T\dot{S}_{tot}=T(\dot{S}_h+\dot{S}_{ph})=4\pi
l^2u(\rho+p)(2\sqrt{M}+\pi lp).
\end{equation}
From the above equation, note that $u<0$ and $\rho+p<0$ the GSL
holds provided $2\sqrt{M}+\pi lp>0$ which implies
\begin{equation}
p\geq-\frac{2\sqrt{M}}{\pi l}.
\end{equation}
Since the pressure of the phantom energy is negative ($p<0$),
therefore the GSL gives us the lower bound on the pressure of the
phantom energy.
\begin{equation}
-\frac{2\sqrt{M}}{\pi l}\leq p<0.
\end{equation}
The GSL in the phantom energy accretion holds within the inequality
(29). Otherwise GSL does not hold which forbid evaporation of BTZ
black hole by the phantom accretion \cite{pavon}. In addition, it is
not clear whether the GSL should be valid in presence of the phantom
fluid not respecting the dominant energy condition \cite{pavon}.

\section{Conclusion}

In this paper, we have investigated the accretion of exotic phantom
energy onto a BTZ black hole. The motivation behind this work is to
study the accretion dynamics in low dimensional gravity. Our
analysis has shown that evolution of mass of a BTZ black hole would
be independent of its mass and will be dependent only on the energy
density and pressure of the phantom energy in its vicinity. Due to
spherical symmetry, the accretion process is simple since the
phantom energy falls radially on the black hole. The accretion would
be much more interesting when additional parameters like charge and
angular momentum are also incorporated in the BTZ spacetime.
Similarly, it would be of much interest to perform the above
analysis in higher ($n+1$) dimensional black hole spacetimes.

We also discussed GSL in the BTZ black hole spacetime. We assumed
that the event horizon of BTZ black hole acts as a boundary of the
thermal system and the phantom energy crossing the event horizon
will change the mass of the black hole. We assumed that the horizon
temperature is in local equilibrium with the temperature of the
matter energy at the event horizon. Under these constraints it is
shown that the GSL holds provided the pressure of the phantom energy
$p$ has an lower bound $p\geq-\frac{2\sqrt{M}}{\pi l}$, on the black
hole parameters ($M$ and $l$).

\subsubsection*{Acknowledgment}

We would like to thank NUST for providing us financial support to
visit ICRANet, Pescara, Italy to present this paper at the Second
Joint Italian-Pakistani Workshop on Relativistic Astrophysics. We
would also thank anonymous referees for their useful comments on
this work. Also MJ would thank Emmanuel N. Saridakis, H. Mohseni
Sadjadi, Diego Pavon and Ahmad Sheykhi for enlightening discussions
during this work.

\end{document}